\documentclass{PoS}
\usepackage[caption=false]{subfig}
\usepackage{graphicx}                                
\usepackage{amsmath,amsfonts,bbm}
\usepackage{amssymb}
\usepackage{mathrsfs}

\newcommand{\pslash}{p\hspace{-2mm}\slash}
\newcommand{\vecpslash}{\vec p\hspace{-2mm}\slash}

\title{Lattice Coulomb propagators, effective energy and confinement}

\ShortTitle{Lattice Coulomb propagators, effective energy and confinement}

\author{\speaker{Giuseppe Burgio}\\
        Institut f\"ur Theoretische Physik\\
        Auf der Morgenstelle 14\\
        72076 T\"ubingen\\
        Germany\\
        E-mail: \email{giuseppe.burgio@uni-tuebingen.de}}
\author{Markus Quandt\\
        Institut f\"ur Theoretische Physik\\
        Auf der Morgenstelle 14\\
        72076 T\"ubingen\\
        Germany\\
        E-mail: \email{markus.quandt@uni-tuebingen.de}}
\author{Hugo Reinhardt\\
        Institut f\"ur Theoretische Physik\\
        Auf der Morgenstelle 14\\
        72076 T\"ubingen\\
        Germany\\
        E-mail: \email{hugo.reinhardt@uni-tuebingen.de}}
\author{Mario Schr\"ock\\
        Institut f\"ur Physik, FB Theoretische Physik\\
        Universit\"at Graz\\
        8010 Graz\\
        Austria\\
        E-mail: \email{mario.schroeck@uni-graz.at}}

\abstract{We show that in the lattice Hamiltonian limit all Coulomb gauge 
propagators are consistent with the Gribov-Zwanziger confinement mechanism, 
with an IR enhanced effective energy for quarks and gluons and a diverging
ghost form factor compatible with a dual-superconducting vacuum. 
Multiplicative renormalizability is ensured for all static correlators, while 
for non-static ones their energy dependence plays a crucial role in this 
respect. Moreover, from the Coulomb potential we can extract the Coulomb 
string tension $\sigma_C \sim 2 \sigma$.}

\FullConference{Xth Quark Confinement and the Hadron Spectrum\\
                 8-12 October 2012\\
		 TUM Campus Garching, Munich, Germany}

\begin{document}

\section{Introduction}

QCD in Coulomb gauge, being best suited to examine the Gribov-Zwanziger (GZ)
confinement ideas \cite{Gribov:1977wm,Zwanziger:1995cv}, has been the subject 
of intense research in the last few years. In a series of papers 
\cite{Burgio:2008jr,Burgio:2009xp,Quandt:2010yq,Reinhardt:2011fq,%
Burgio:2012ph,Burgio:2012bk}, which we will briefly summarize here, we have 
analyzed the behaviour of all relevant two-point functions on the lattice and 
compared them with the corresponding predictions of Hamiltonian variational 
calculations \cite{Szczepaniak:2001rg,Feuchter:2004mk,Epple:2006hv}, 
concentrating on the features relevant for the GZ scenario.

As Gribov in his seminal paper noticed, for non-Abelian theories most 
gauge conditions admit several solutions and the corresponding Faddeev-Popov 
(FP) mechanism is not sufficient to define the functional integral 
beyond perturbation theory. The field-configuration space must therefore be 
restricted to a domain, continuously connected to the origin, where the gauge 
condition at hand always possesses unique solutions. He then showed how, as 
soon as the fields cross the boundary of such region, the ghost dressing 
function can acquire a singularity; the ``no-pole'' condition for the FP-ghost 
at non-vanishing momentum is then necessary to implement the restriction. 
In particular, in Coulomb 
gauge, he argued how such restriction can imply a diverging gluon self-energy, 
motivating its disappearance from the physical spectrum. 

Many issues remain of course in the above description open. Gribov based his 
conjectures on more or less heuristic arguments. Zwanziger later tried to 
put the whole set-up on a more solid basis, while variational calculations, 
which are viable in Coulomb gauge since they by-pass the explicit
construction of the gauge invariant Hilbert space \cite{Burgio:1999tg}, did 
provide some insight on the relation of the GZ-mechanism to the Hamiltonian 
formulation. In both cases, however, approximations need to be made; although 
many authors tackled the problems during the years \cite{Cucchieri:2000kw,%
Langfeld:2004qs,Nakagawa:2007fa,Voigt:2008rr,Nakagawa:2008zza,%
Nakagawa:2008zzc}, a satisfactory non-perturbative cross-check from lattice 
calculation was hindered for a long time by the presence of strong 
discretization effects. We have shown \cite{Burgio:2008jr,%
Burgio:2012ph,Burgio:2012bk} how for each propagator a mixture of improved 
actions and separate treatment of their energy dependence can quite 
effectively solve such problems, allowing an explicit check of the GZ-scenario.

\begin{figure}[htb]
\subfloat[][]{\includegraphics[width=0.49\textwidth,height=0.44\textwidth]{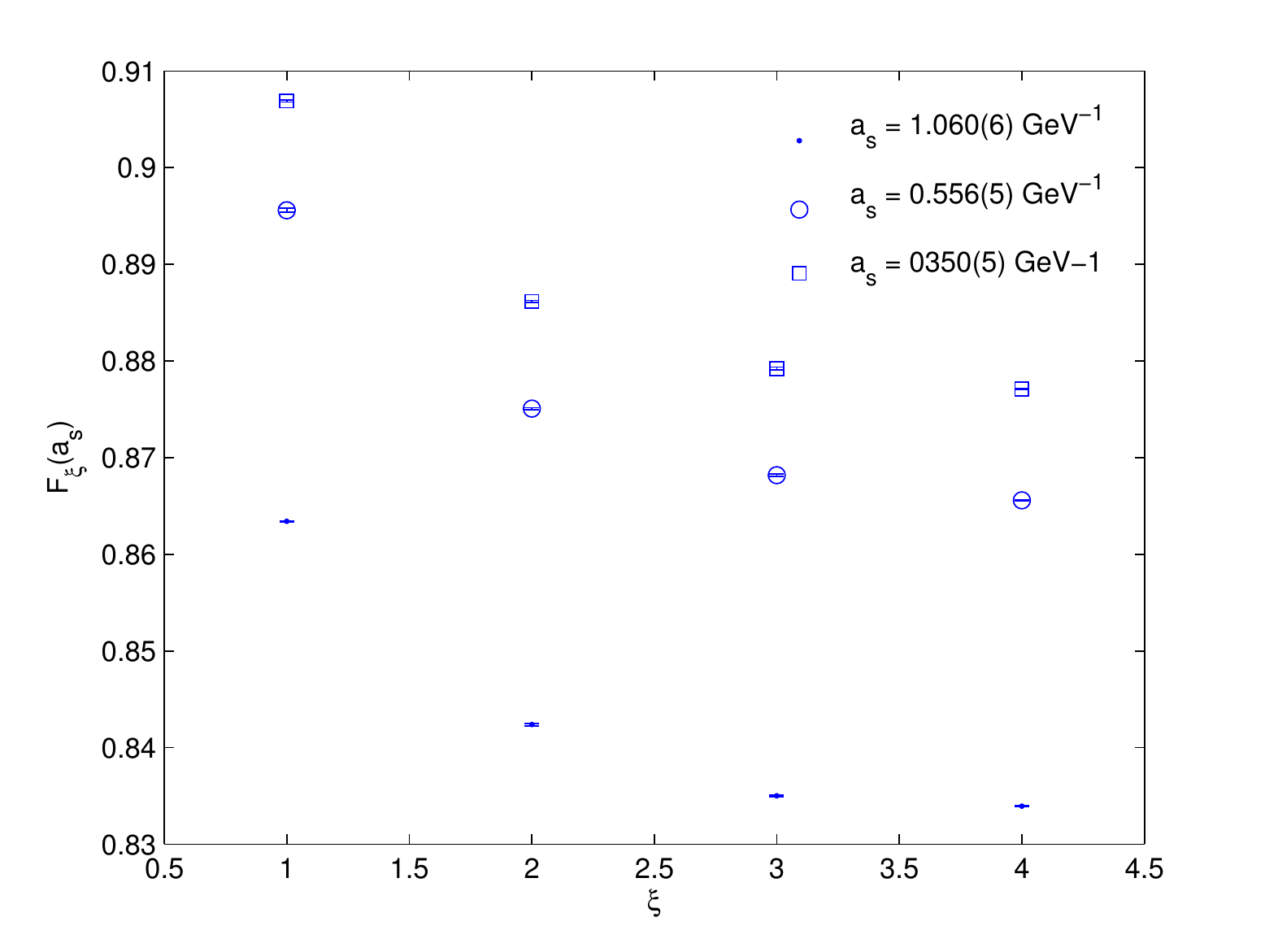}}
\subfloat[][]{\includegraphics[width=0.49\textwidth,height=0.44\textwidth]{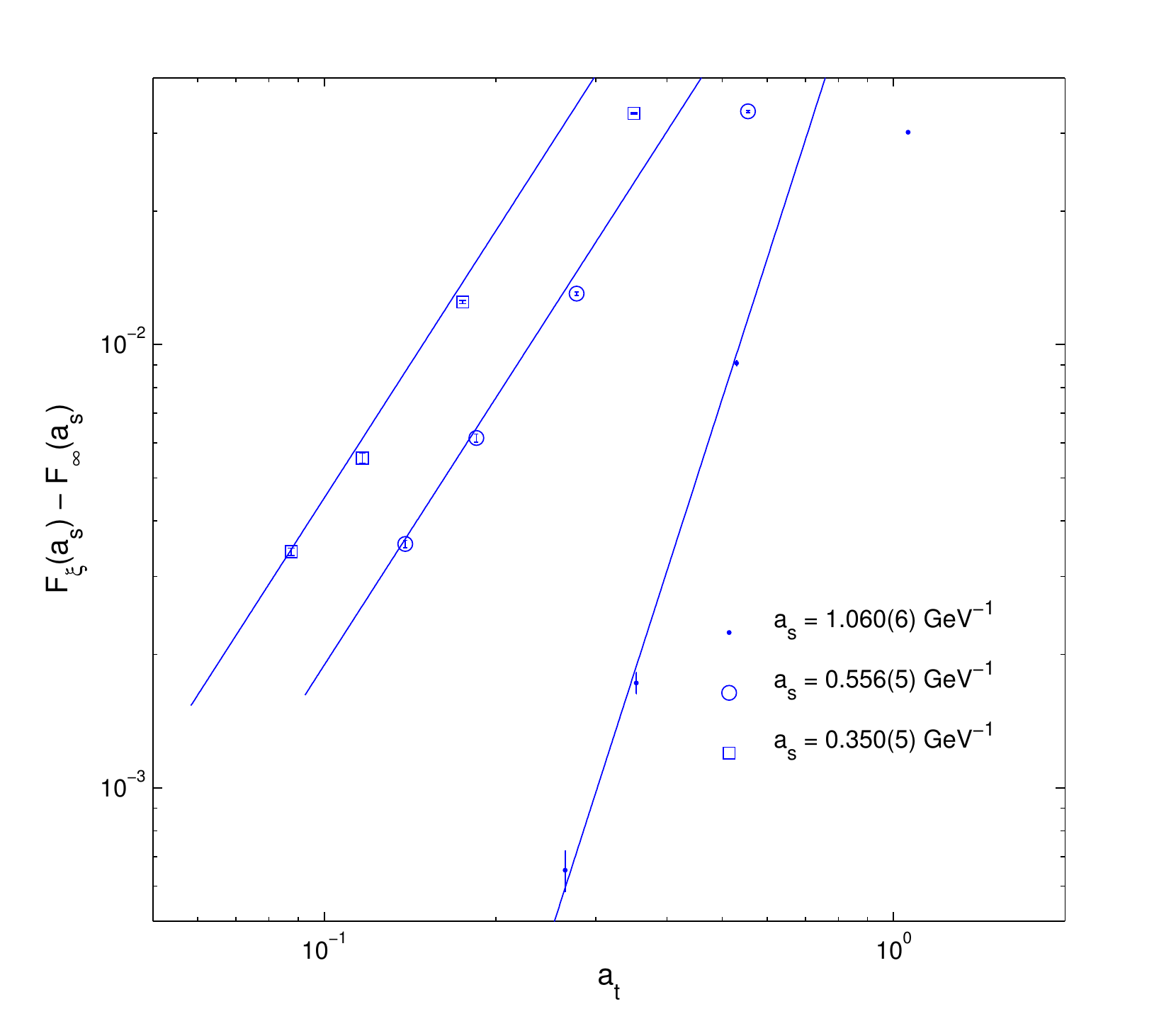}}
\caption{{(a)}: dependence of the gauge fixing functional $F_\xi(a_s)$ 
on the anisotropy $\xi = a_s/a_t$ at fixed spatial cut-off $a_s$. {(b)}: 
Deviation of the gauge fixing functional from the Hamiltonian limit,
$F_\xi(a_s) - F_{\infty}(a_s)$, as a function of the temporal lattice spacing $a_t$,
together with its leading power corrections.}
\label{fig1}
\end{figure}

As first suggested in \cite{Burgio:2008jr}, the size of discretization effects 
can be investigated on anisotropic lattices, where the time and space like 
cut-off $a_t$, $a_s$ are kept different. In Fig.~\ref{fig1} we show the effect 
of taking the limit $a_t \to 0$, which controls the lattice Hamiltonian 
limit, on the SU(2) Coulomb gauge functional calculated at fixed space-like 
cut-off, i.e. RG-point. 
Perturbative determinations of the latter \cite{Burgio:1996ji,Burgio:2003in} 
are not precise enough and we have re-checked the only non-perturbative 
calculation found in the literature \cite{Ishiguro:2001jd}; details can be
found in \cite{Burgio:2012bk}, as well as the details of the gauge fixing
algorithm, which adapts those introduced in 
\cite{Bogolubsky:2005wf,Bogolubsky:2007bw}.
Following the ideas in \cite{Burgio:2008jr}, a 
first anisotropic analysis in SU(3) had been attempted in 
\cite{Nakagawa:2011ar}.

From the continuum analysis and from our results in 
\cite{Burgio:2008jr,Burgio:2009xp} we know that in the pure gauge
sector the static gluon propagator, the static Coulomb potential and the 
ghost form factor should obey:
\begin{equation}
\begin{array}{l}
D(\vec{p})\; = \frac{\displaystyle |\vec{p}|}{\displaystyle 
\sqrt{|\vec{p}|^4+M^4}}\\
\\
V_C(\vec{p})  = \frac{\displaystyle 8 \pi \sigma_C}{\displaystyle |{\vec{p}}|^4}
 + \frac{\displaystyle \eta}{\displaystyle |{\vec{p}}|^2}
      + \mathcal{O}(1)
\end{array}
\qquad
d(\vec{p}) \simeq \left\{\begin{array}{c l}
\frac{\displaystyle 1}{\displaystyle |\vec{p}|^{\kappa_{\rm gh}}}&\quad|\vec{p}|\ll \Lambda\\ 
&\\
\frac{\displaystyle 1}{\displaystyle \log^{\gamma_{\rm gh}}{\frac{|\vec{p}|}{m}}} &\quad|\vec{p}|\gg \Lambda
\end{array}\right.
\label{eq_1_1}
\end{equation}
where the Gribov mass $M\simeq 1$~GeV and for the gluon self-energy 
$\omega_A = D^{-1}(\vec{p})$ holds. 
The quark propagator, the fermion self energy and the 
running mass $M(|\vec{p}|)$ take the form
\cite{Burgio:2012ph}:
\begin{equation}
\begin{array}{c}
 S(\vec{p},p_4) = \frac{\displaystyle Z(\vec{p})}{\displaystyle i\vecpslash 
      + i\pslash_4 \alpha(\vec{p})+M(\vec{p})}\qquad
\omega_F(|\vec{p}|) =
    \frac{\displaystyle \alpha(|\vec{p}|)}{\displaystyle Z^2(|\vec{p}|)}\sqrt{\vec{p}^2 + 
      M^2(|\vec{p}|)}\\
\\
M(|\vec{p}|) = \frac{\displaystyle m_\chi(m_b)}{\displaystyle 1+b\, \frac{|\vec{p}|^2}{\Lambda^2}\,
\log{\left(e+\frac{|\vec{p}|^2}{\Lambda^2}\right)}^{-\gamma}}+
\frac{\displaystyle m_r(m_b)}{\displaystyle \log{\left(e+\frac{|\vec{p}|^2}{\Lambda^2}\right)}^{\gamma}}\,,
\end{array}
\label{eq_1_2}
\end{equation}
where $Z$ is the field renormalization function, $\alpha$ the energy
renormalization function, $m_b$ the bare quark mass, $m_\chi(m_b)$ the chiral mass
and $ m_r(m_b)$ the renormalized running mass \cite{Burgio:2012ph}. In the 
following we shall verify such behaviour and determine the relevant parameters.

\section{Results}

\subsection{Ghost form factor}

A careful analysis of the ghost form factor in the Hamiltonian limit $a_t \to 0$ 
shows that its UV behaviour agrees with Eq.~(\ref{eq_1_1}), with 
$\gamma_{\rm gh} =1/2$, confirming continuum predictions, and $m = 0.21(1)$~GeV, 
see Fig.~\ref{fig1}~(a). In the IR going to higher anisotropies increases
the exponent $\kappa_{\rm gh}$, as shown in Fig.~\ref{fig1}~(b), where we plot
$|\vec{p}|^{\kappa_m} \, d(\vec{p})$, with $\kappa_m$ the IR exponent
for $\xi=1$, as a function \begin{figure}[htb]
\subfloat[][]{\includegraphics[width=0.49\textwidth,height=0.44\textwidth]{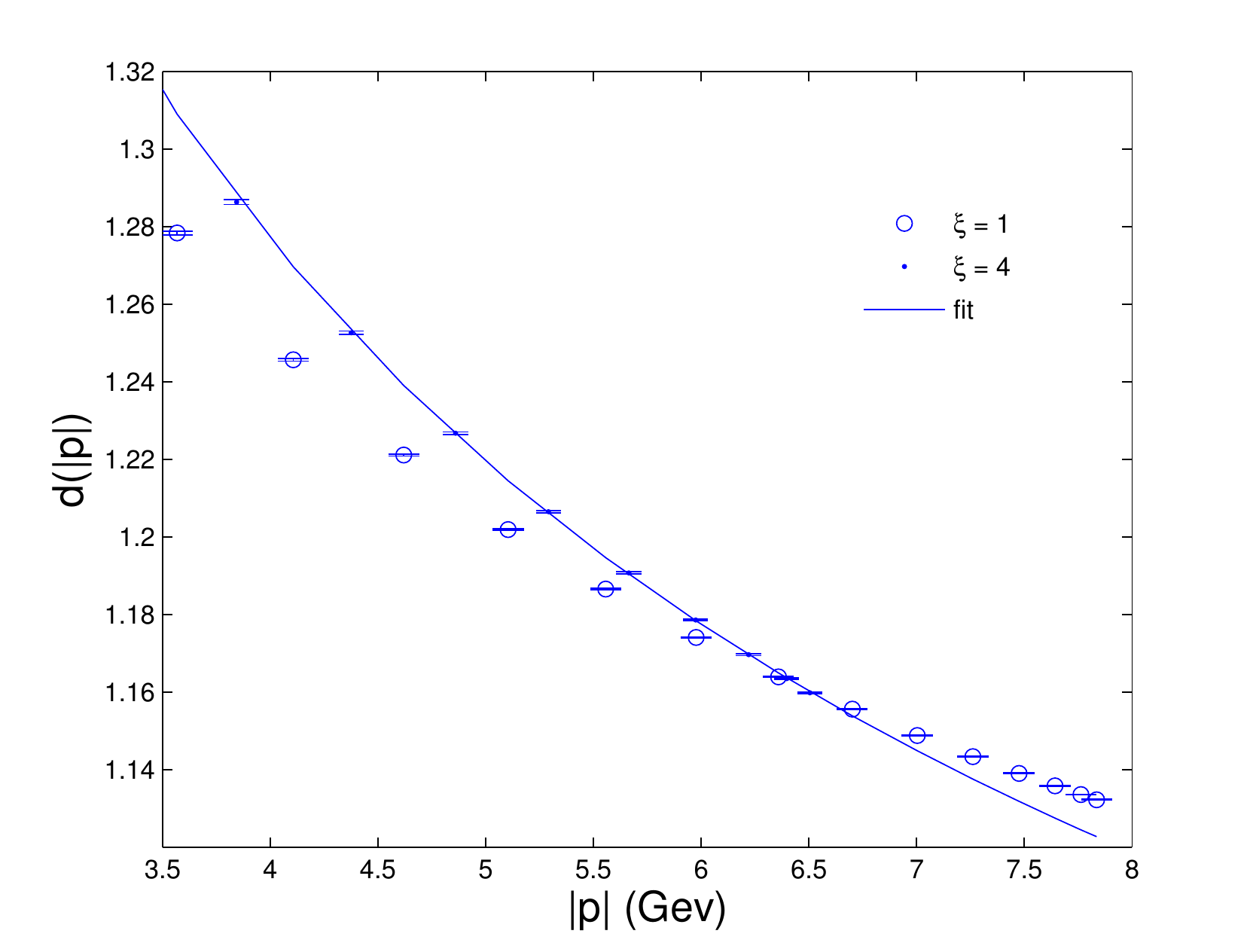}}
\subfloat[][]{\includegraphics[width=0.49\textwidth,height=0.44\textwidth]{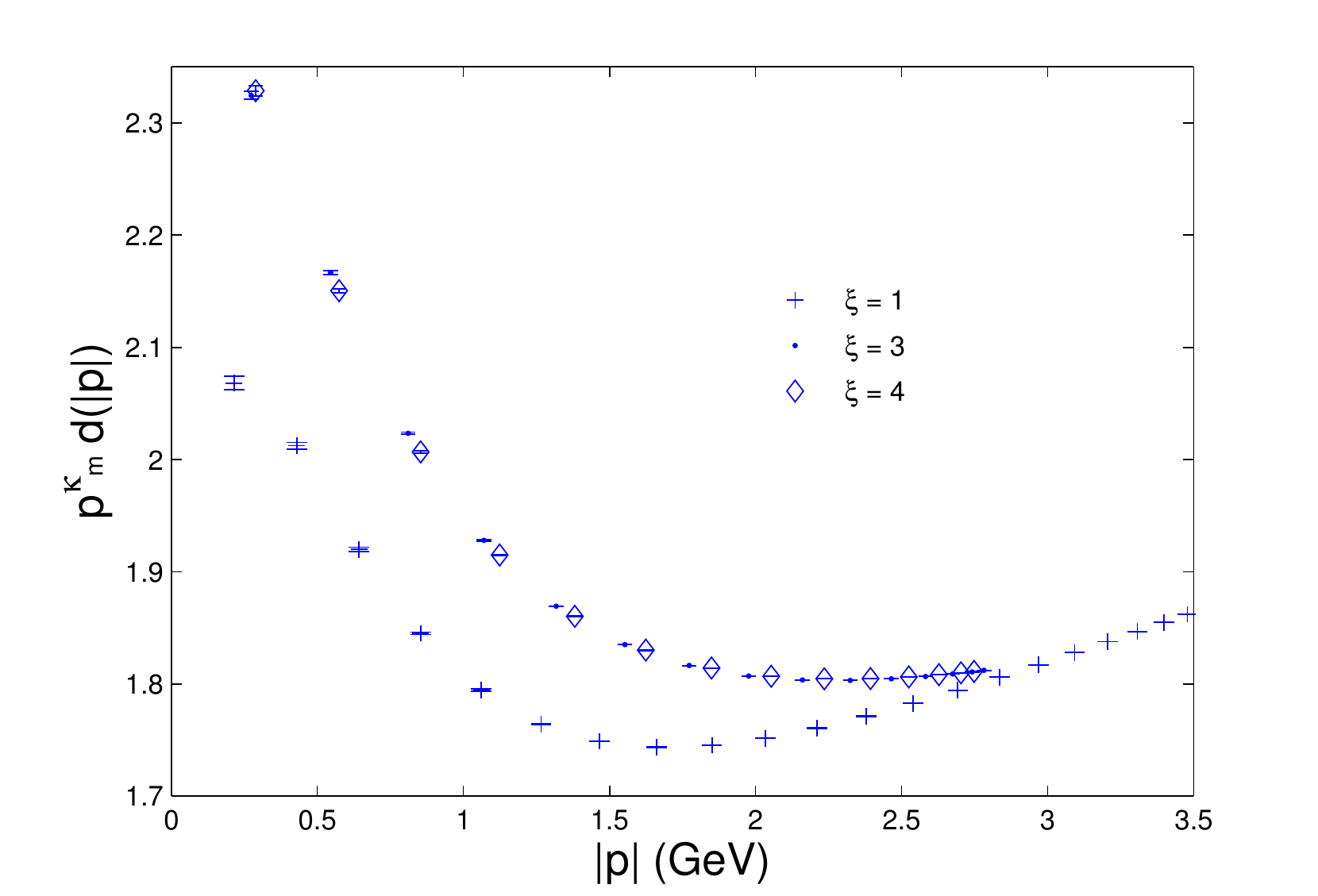}}
\caption{(a): UV behavior of $ d({\vec{p}})$ compared with 
Eq.~(\protect\ref{eq_1_1}). (b): IR behavior of 
$|{\vec{p}}|^{\kappa_m}\, d({\vec{p}})$, both for different anisotropies $\xi$.}
\label{fig2}
\end{figure}
of the anisotropy. The limit $\xi\to\infty$ gives
$\kappa_{\rm gh} \gtrsim 0.5$, confirming the GZ-scenario. This however 
disagrees with some continuum predictions $\kappa_{\rm gh} =1$, deriving from 
the assumption of the finiteness of the static ghost-gluon vertex. Whether 
this is indeed correct and algorithmic improvements could change the lattice 
result is still a matter of investigation.

\subsection{Coulomb potential}

\begin{figure}[htb]
\subfloat[][]{\includegraphics[width=0.49\textwidth,height=0.44\textwidth]{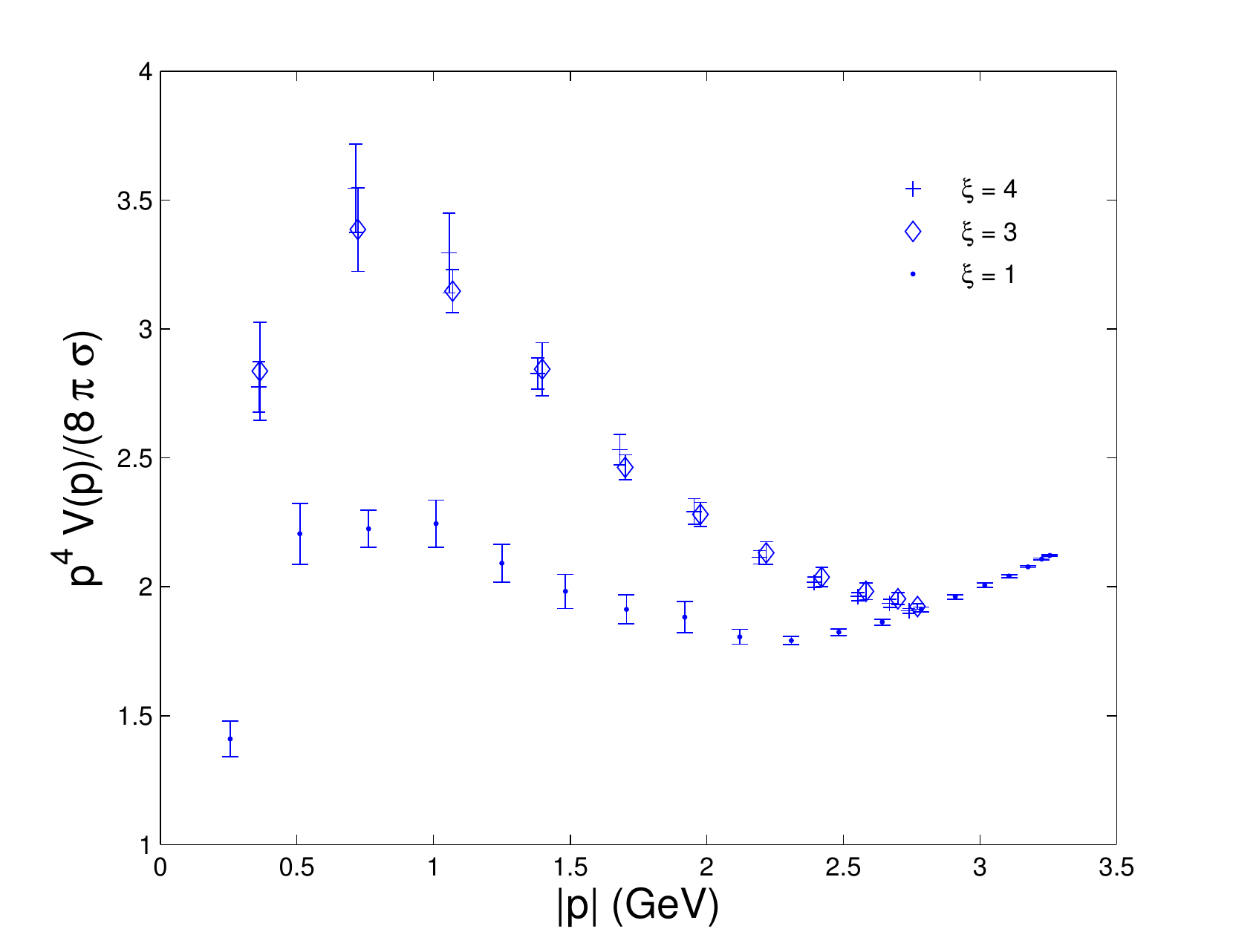}}
\subfloat[][]{\includegraphics[width=0.48\textwidth,height=0.43\textwidth]{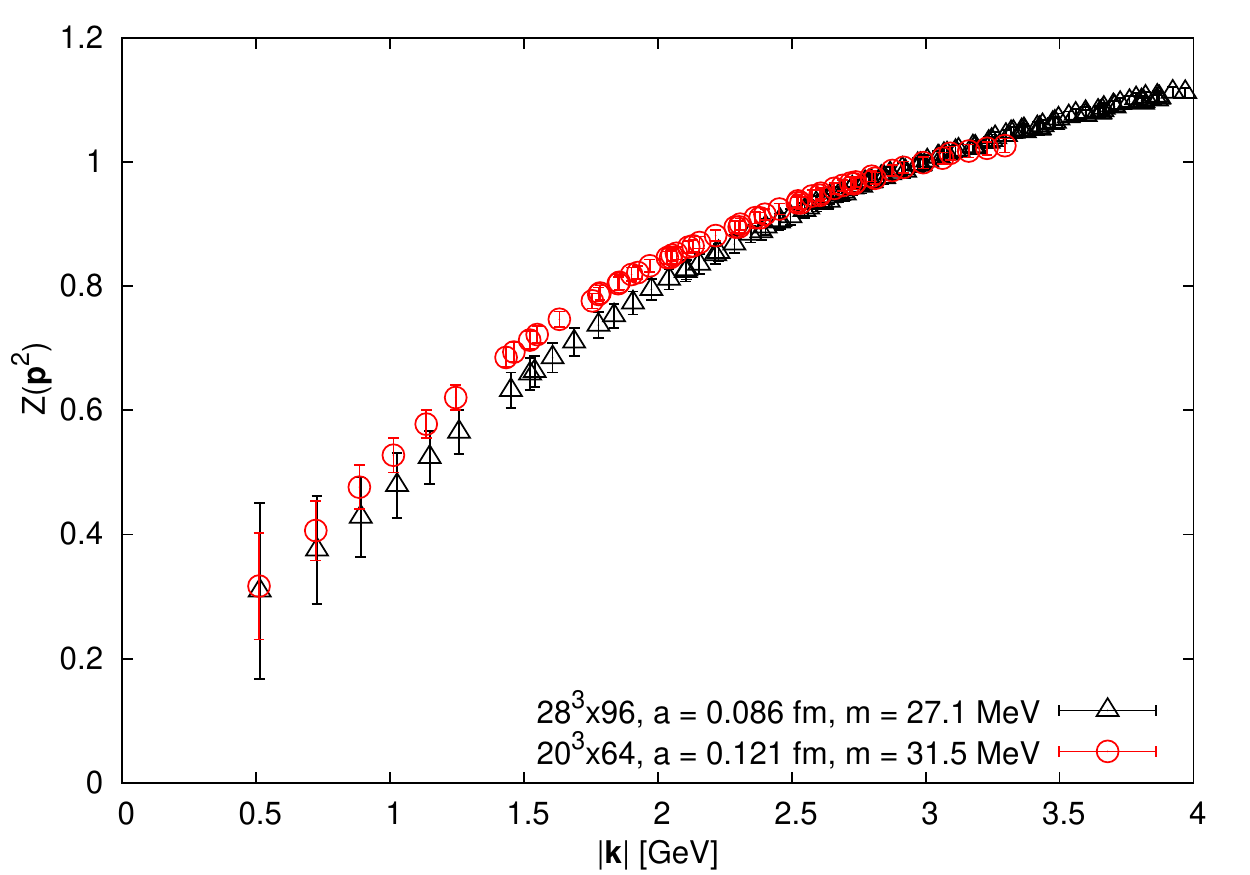}}
\caption{{(a)}: Infrared behavior of $|{\vec{p}}|^4\, V_C(\vec{p})/(8 \pi \sigma)$ for 
different anisotropies $\xi$. {(b)}: Quark field renormalization
function $Z(|{\vec{p}}|)$.}
\label{fig3}
\end{figure}

In Fig.~\ref{fig3}~(a) we show $|{\vec{p}}|^4 V_C(|{\vec{p}}|)$ as obtained 
from different anisotropies. Fitting the results to Eq.~(\ref{eq_1_1}) we get,
in the Hamiltonian limit $\xi\to\infty$ $\sigma_C = 2.2(2) \,\sigma$, as
expected from Zwanziger's predictions \cite{Zwanziger:2002sh}. 

\subsection{Quark propagator}

Our calculations were all made on a set of configurations generated by the MILC
collaboration \cite{Bazavov:2009bb}, see \cite{Burgio:2012ph} for details.
The use of improved actions is crucial to establish the scaling
properties of the Coulomb gauge quark propagators. This is very
similar to the situation in Landau gauge, see e.g. 
\cite{Bowman:2002bm,Bowman:2005vx,Parappilly:2005ei}, whose techniques
we have adapted to our case. 

\begin{figure}[htb]
\subfloat[][]{\includegraphics[width=0.49\textwidth,height=0.44\textwidth]{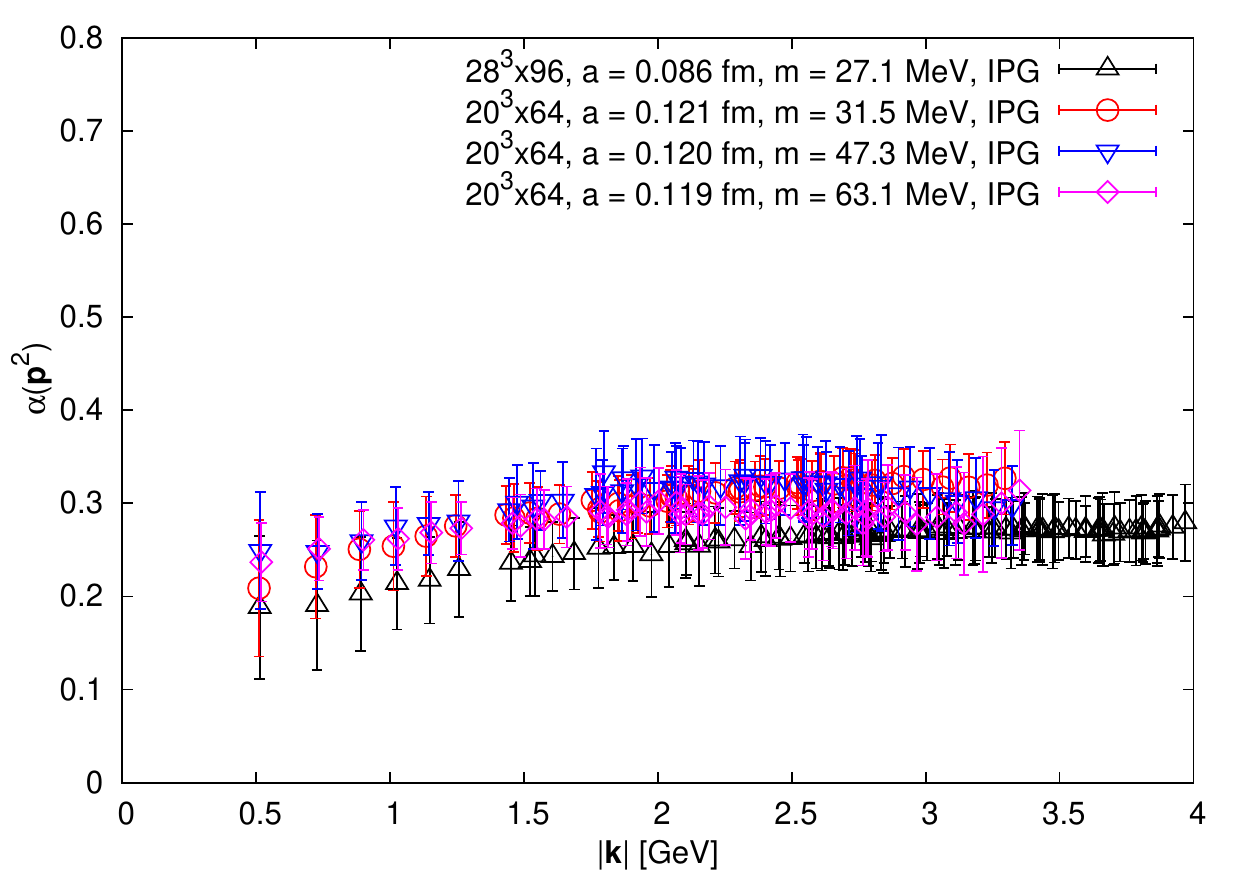}}
\subfloat[][]{\includegraphics[width=0.49\textwidth,height=0.44\textwidth]{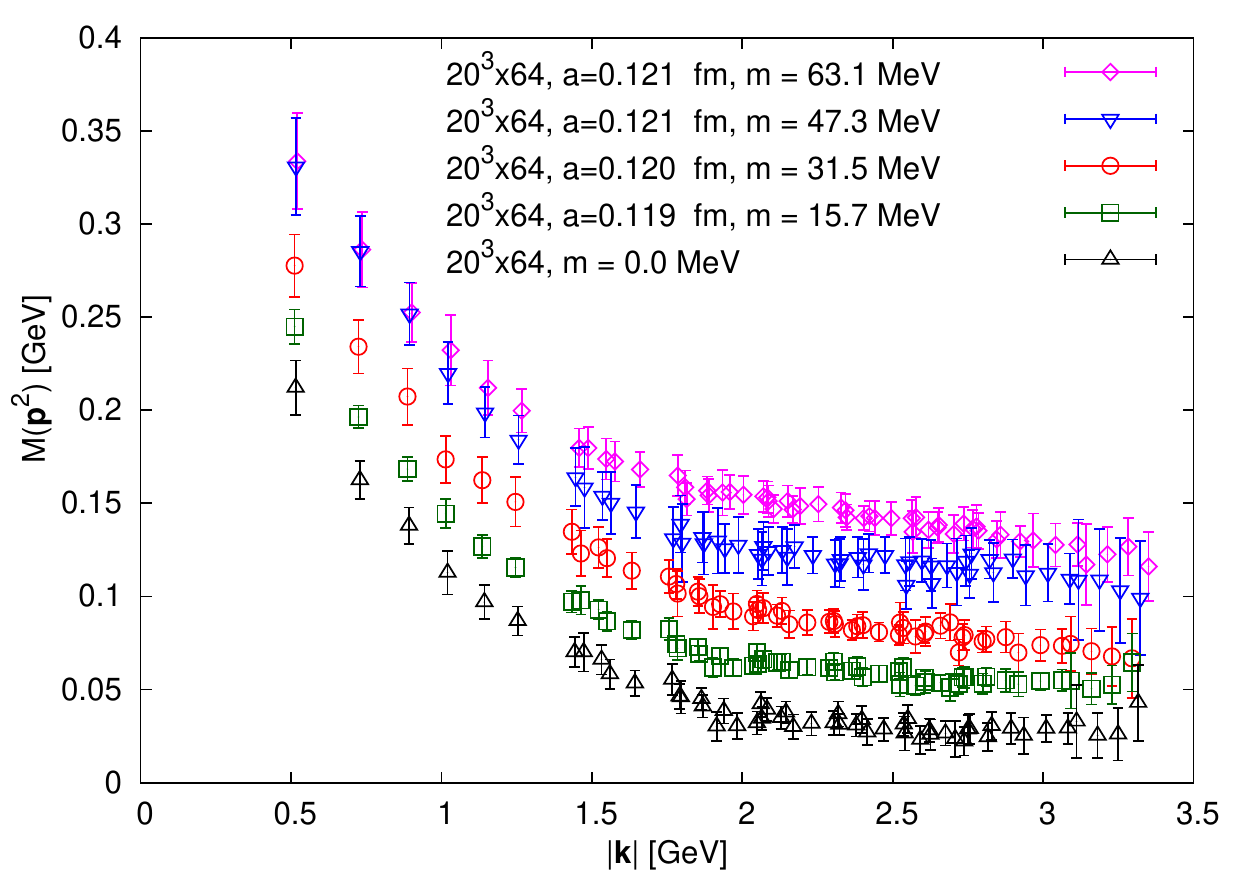}}
\caption{{(a)}: Energy renormalization function $\alpha(|{\vec{p}}|)$. {(b)}: 
Running mass $M(|{\vec{p}}|)$.}
\label{fig4}
\end{figure}

Fig.~\ref{fig3}~(b) shows the scaling of the renormalization function 
$Z(|{\vec{p}}|)$ for configurations calculated at similar bare quark mass, 
while the RG-invariant functions $\alpha(|{\vec{p}}|)$ and $M(|{\vec{p}}|)$
are given in Fig.~\ref{fig4}. Their behaviour agrees with theoretical
expectations, see Eq.~(\ref{eq_1_2}).

\begin{figure}[htb]
\subfloat[][]{\includegraphics[width=0.49\textwidth,height=0.44\textwidth]{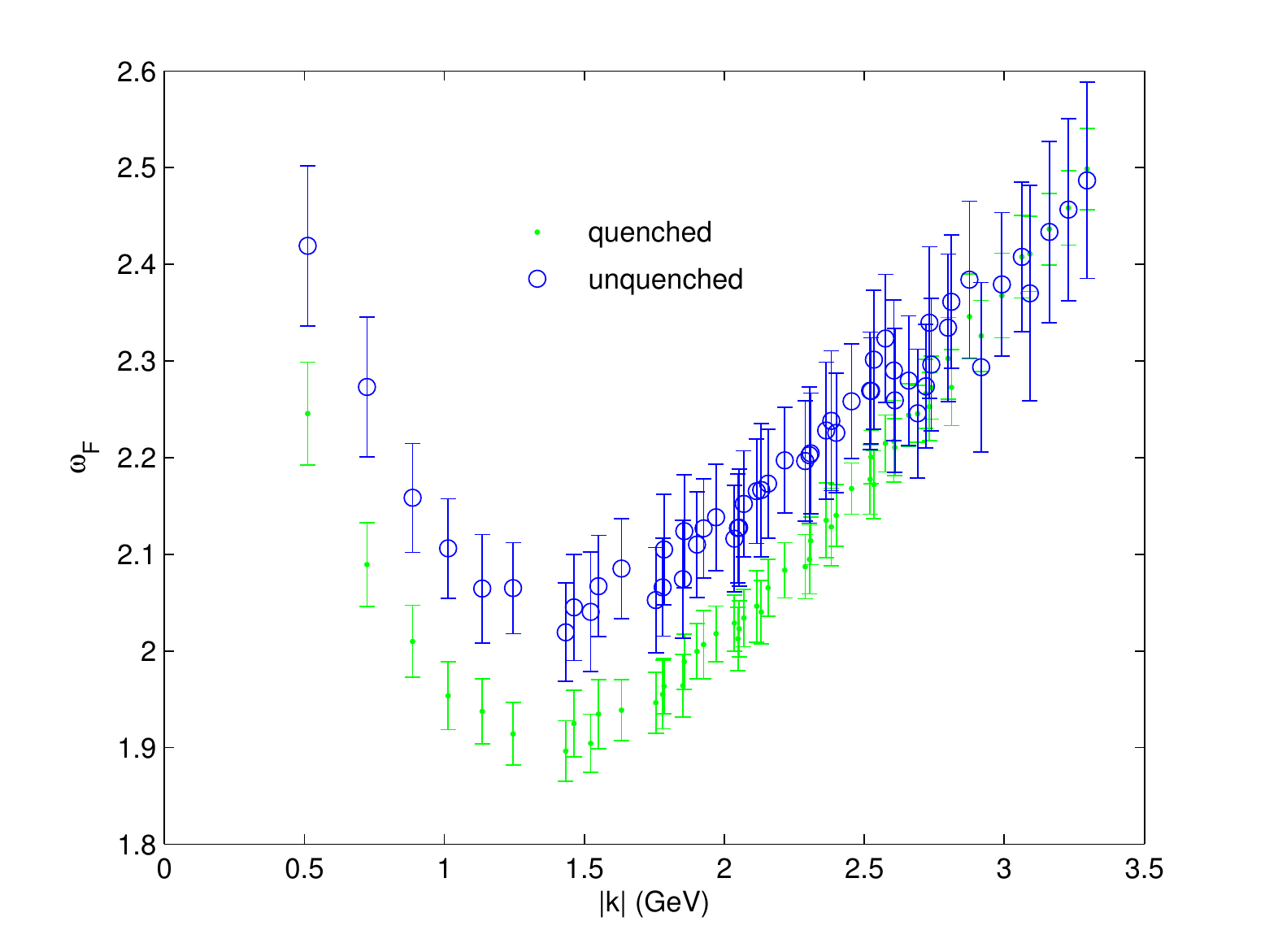}}
\subfloat[][]{\includegraphics[width=0.49\textwidth,height=0.44\textwidth]{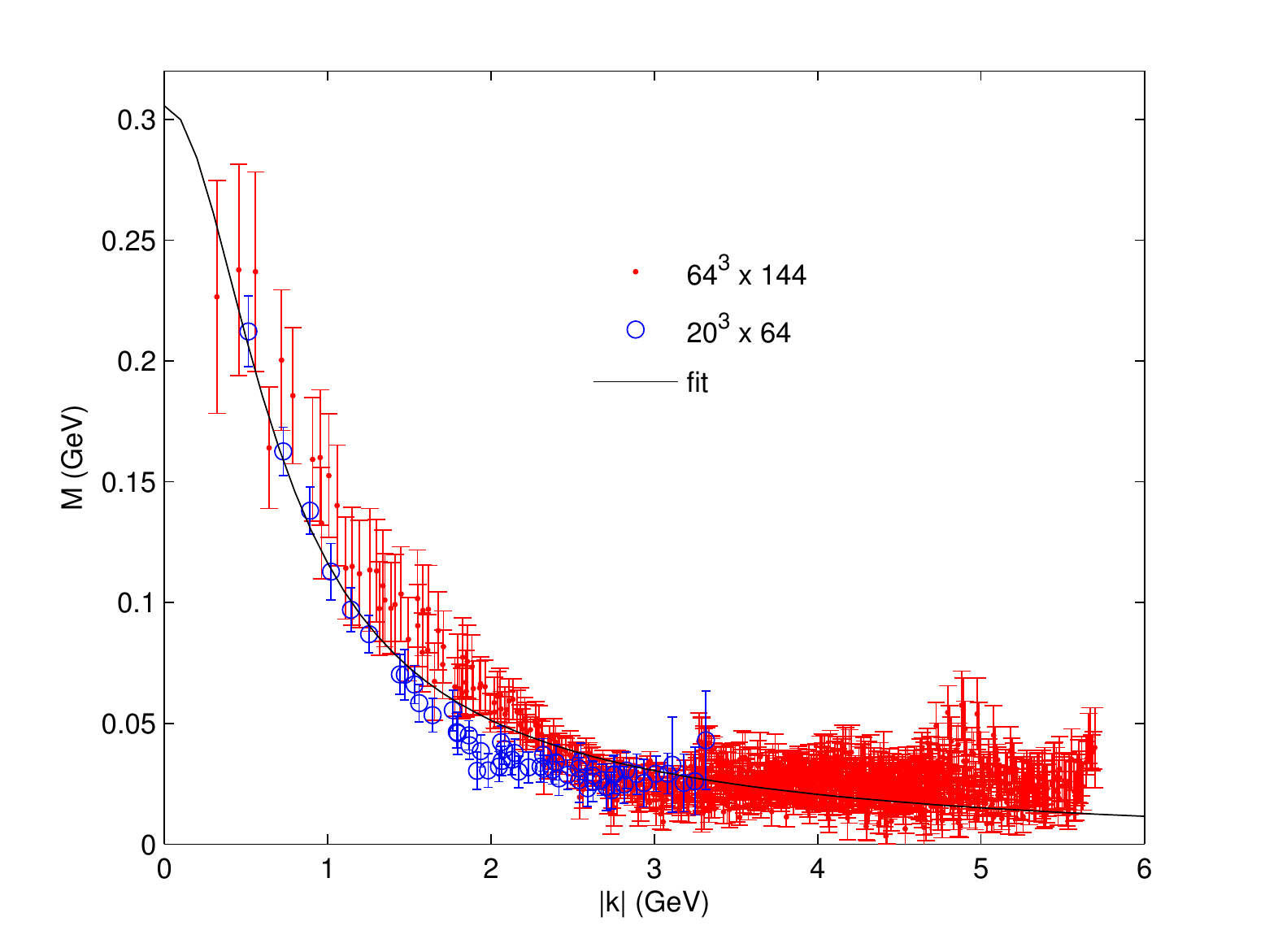}}
\caption{{(a)}: Quark self energy $\omega_F(|\vec{p}|)$. {(b)}: Running
mass $M(|{\vec{p}}|)$ in the chiral limit $m_b\to 0$; see
Eq.~(\protect\ref{eq_1_2}).}
\label{fig5}
\end{figure}

Our most interesting results are given in Fig.~\ref{fig5}. Analogously to
the gluon self-energy $\omega_A(|\vec{p}|)$, the quark self energy 
$\omega_F(|\vec{p}|)$ has a turn-over at $|\vec{p}| \sim 1$~GeV, 
clearly departing from the behaviour of a free particle, and diverging in the 
IR, see Fig.~\ref{fig5}~(a); although awaiting confirmation on larger 
lattices, this would in principle extend the Gribov argument for its 
disappearance from the physical spectrum to full QCD. Moreover, as 
Fig.~\ref{fig5}~(b) shows, the running mass $M(|{\vec{p}}|)$ we obtain is 
quantitatively compatible with our phenomenological expectations
from chiral symmetry breaking. Fitting it to Eq.~(\ref{eq_1_2})
we obtain $b = 2.9(1)$, $\gamma = 0.84(2)$, $\Lambda = 1.22(6)$~GeV, 
$m_\chi(0) = 0.31(1)$~GeV, with $\chi^2$/d.o.f.$=1.06$. 

\section{Conclusions}

We have shown that the GZ confinement scenario is realized
in Coulomb gauge. The ghost form factor $d(|\vec{p}|)$ is IR divergent
with an exponent $\kappa_{\rm gh} \gtrsim 0.5$, which implies Gribov's no-pole 
condition and a
dual-superconducting scenario \cite{Reinhardt:2008ek}; 
the gluon propagator satisfies
the Gribov formula, implying an IR diverging self-energy, and the Coulomb 
string tension is roughly twice the physical string tension. Moreover from the 
quark propagator we can easily extract the quark self energy 
$\omega_F(|\vec{p}|)$, which is also compatible with an IR divergent behaviour, 
and the running mass $M(|{\vec{p}}|)$, which gives a constituent quark mass
of $m_\chi(0) = 0.31(1)$~GeV.

This is in contrast to Landau gauge, where BRST symmetry seems
to be non-perturbatively broken, violating the Kugo-Ojima
confinement scenario \cite{Kugo:1979gm}, while the GZ confinement scenario 
cannot be realized without the explicit introduction of an horizon function, 
see e.g. \cite{Vandersickel:2012tz} for a recent review; its physical
implications and
how these can be related to the presence of
dim-2 condensates \cite{Burgio:1997hc,Akhoury:1997by,Boucaud:2000ey}
are an interesting issue still debated in the literature 
\cite{Cucchieri:2011ig}.


\end{document}